\documentstyle[amssymb,aps,prl,preprint,epsf]{revtex}

\begin{document}
\title{A new cosmological tracker solution for Quintessence}
\author{L. Arturo Ure\~{n}a-L\'{o}pez\thanks{%
E-mail: lurena@fis.cinvestav.mx} and Tonatiuh Matos\thanks{%
E-mail: tmatos@fis.cinvestav.mx}}
\address{Departamento de F\'{\i}sica, \\
Centro de Investigaci\'on y de Estudios Avanzados del IPN,\\
AP 14-740, 07000 M\'exico D.F., MEXICO.\\
}
\date{\today}
\maketitle

\begin{abstract}
In this paper we propose a quintessence model with the potential $V(\Phi
)=V_{o}\left[ \sinh {(\alpha \,\sqrt{\kappa _{o}}\Delta \Phi })\right]
^{\beta }$, which asymptotic behavior corresponds to an inverse power-law
potential at early times and to an exponential one at late times. We
demonstrate that this is a tracker solution and that it could have driven
the Universe into its current inflationary stage. The exact solutions and
the description for a complete evolution of the Universe are also given. We
compare such model with the current cosmological observations.
\end{abstract}

\draft

\narrowtext

Recent observations of Ia supernovae revealed the existence of a certain
vacuum energy with a negative equation of state, $\omega <0$, being $%
p=\omega \rho $. It seems that the contribution of this vacuum energy, also
called dark energy, leads the Universe to reach the critical energy density
and to accelerate its expansion\cite{perlmutter}. Furthermore, if these observations can be confirmed, they indicate that this missing energy will dominate the future evolution of the Cosmos. The most reliable models for such missing energy are the cosmological constant ($\Lambda $)\cite{triangle} and a fluctuating, inhomogeneus scalar field, rolling down a scalar potential, called Quintessence ($Q$)\cite{stein,stein2}. For the later case, great effort has been made to determine the appropiate scalar potential that could explain current cosmological observations\cite{stein,stein2,ferr,barr,silviu}. One example, is the pure exponential potential\cite{ferr,barr}. It has the advantage that it mimics the dominant density background, but nucleosynthesis constraints require that the scalar field contribution be $\Omega _{\Phi }\leq 0.2$, which indicates that the
scalar field would never dominate the Universe\cite{ferr}. On the other
hand, a special group of scalar potentials has been proposed in order to
have $\rho _{\Phi }\sim \rho _{M}$ at the present epoch: the tracker
solutions\cite{stein2}. In these solutions, the cosmology at late times is
extremely insensitive to initial conditions, reducing the fine tuning
problem. One of these potentials is the pure inverse power-law one, $V(\Phi
)\sim \Phi ^{-\alpha }$\cite{stein2,silviu,peebles}. Althought it reduces
the fine tuning and the cosmic coincidence problem, the predicted value for
the current equation of state for the quintessence is not in good agreement
with supenovae results\cite{stein2}. The same problem arises with the
inverse power-law-like potentials. Another example are the new potentials
proposed in\cite{varun}. They avoid efficientely the troubles stated above,
but it is not possible to determine unambigously their three parameters. 

In a previous work, we showed that if the Universe is completely dominated
by the scalar energy density with a constant equation of state, then it
naturally arises that the effective scalar field potential is an exponential
one \cite{urena}. In this letter, we combine this fact of the late Universe
with a power-law-like potential , which seems to be a good candidate for the
early Universe. As we will see, this combination avoids the problems of the
exponential potential and coincides very well with observations. In order to
do so, we propose a new cosmological tracker solution with the scalar field
potential

\begin{equation}
V(\Phi )=V_{o}\left[ \sinh {\left(\alpha \,\sqrt{\kappa _{o}}\Delta \Phi \right)}\right]
^{\beta },  \label{potential}
\end{equation}

\noindent which asymptotic behavior corresponds to an inverse power-law-like
potential at early times and to an exponential one at late times. We will
analyze the exact solutions of the evolution equations for this case and we
will determine all the parameters in order to have a tracker solution and an
accelerated expansion at the present time.

We start with the flat FRW metric

\begin{equation}
ds^{2}=-dt^{2}+a^{2}(t)\left[ dr^{2}+r^{2}\left( d\theta ^{2}+\sin
^{2}(\theta )d\phi ^{2}\right) \right].
\end{equation}

\noindent Thus, the equation of evolution for the scalar field $\Phi $ is

\begin{equation}
\ddot{\Phi}+3\frac{\dot{a}}{a}\dot{\Phi}+\frac{dV}{d\Phi }=0,  \label{cphi}
\end{equation}

\noindent and the Hubble parameter satisfies the Friedmann equation

\begin{equation}
\left( \frac{\dot{a}}{a}\right) ^{2} =\frac{\kappa _{o}}{3} \left( \rho
+\rho _{\Phi }\right)  \label{fried}
\end{equation}

\noindent where $\rho _{\Phi }=\frac{1}{2}\dot{\Phi}^{2}+V(\Phi )$ is the
density of the scalar field, $\rho $ is the density of the baryons, plus
neutrinos, plus radiation, etc., and $\kappa _{o}=8\pi G=M_{P}^{-2}$. The
equations (\ref{cphi}) and (\ref{fried}) can be written in a more convenient
form if we define the function $F(a)$ such that $V(\Phi (a))=F(a)/a^{6}$\cite
{chimen}. A first integral of the field equation (\ref{cphi}) can be found

\begin{equation}
\frac{1}{2}\dot{\Phi}^{2}+V(\Phi )=\frac{6}{a^{6}}\int da\frac{F}{a}+\frac{C%
}{a^{6}}=\rho _{\Phi },  \label{firint}
\end{equation}

\noindent being $C$ an integration constant. If the scale factor is
considered as the independent variable, the field equations can be
integrated up to quadratures \cite{chimen}

\begin{eqnarray}
\Delta t &=&\sqrt{3}\int {\frac{da}{a\sqrt{\kappa _{0}(\rho _{\Phi }+\rho )}}%
}  \label{time} \\
\Delta\Phi &=&\sqrt{6}\int {\frac{da}{a}}\left[ {\frac{\rho _{\Phi}-F/a^{6}}{%
\rho _{\Phi }+\rho}}\right] ^{1/2},  \label{quad}
\end{eqnarray}

\noindent where we have normalized the scalar field in units of $M_P$. The
exact solution for the potential (\ref{potential}) can be found if we choose 
$F(a)=B a^s$ and $C=0$, where $6-s=3(1+\omega^{\ast})$, being $s$ a constant (we shall denote the quantities in the solution with an asterisk ($^*$)). Then, the scalar energy density scales as $\rho^{\ast}= \rho^{\ast}_o \, a^{-3(1+\omega^{\ast})}$, being $\rho^{\ast}_o = \frac{2\, B}{1-\omega^{\ast}}$. The complete solution is\cite{chimen}

\begin{eqnarray}
\alpha \, \Delta \Phi = arccoth{\left( 1+\frac{\rho_{oD}}{\rho^{\ast}_o}
a^{3(\omega^{\ast}-\omega_D)} \right)^{1/2}}  \label{phi} \\
\alpha = -\frac{3(\omega^{\ast} -\omega_D)}{2 \sqrt{3(1+\omega^{\ast})}}, \label{alpha} \\
\beta = \frac{2 \, (1+ \omega^{\ast})}{\omega^{\ast}-\omega_D}  \label{beta} \\
\rho^{\ast}_o = \left(\frac{2 \, V_o}{1-\omega^{\ast}} \, \rho^{-\beta/2}_{oD}\right)^{\frac{1}{1-\, \beta/2}}  \label{rho} \\
\sqrt{\frac{\kappa_o}{3}} \, \Delta t = \sqrt{\frac{12}{\rho^{\ast}_o}} \, \frac{a^{\frac{3}{2} (1+\omega^{\ast})}}{3(1+\omega^{\ast})} \times  \nonumber \\
_2F_1\left( \frac{1}{2},\frac{\beta}{4},\frac{\beta +4}{4},-\frac{\rho_{oD}}{\rho^{\ast}_o} \, a^{3(\omega^{\ast}-\omega_D)}\right) \\
V^\prime (\Phi) = \alpha \, \beta \, \tanh^{-1}{\left( \alpha \, \Delta \Phi
\right)} \, V(\Phi)  \label{first} \\
V^{\prime \prime} = \left[ (\beta -1) \,\tanh^{-2}{\left( \alpha \, \Delta
\Phi \right)} + 1 \right] \alpha^2 \, \beta \, V(\Phi).  \label{second}
\end{eqnarray}

\noindent where $\rho_D = \rho_{oD} \, a^{-3(1+\omega_D)}$ and $\omega _{D}$ respectively are the energy density and the equation state for the dominant component at each stage of the evolution of the Universe. For given values of $\alpha $ and $\beta $ we will have this solution only when eqs. (\ref{alpha},\ref{beta}) are simultaneously satisfied and $\rho^{\ast} \sim \rho_D$.

As state above, the important feature of this potential is its
asymptotic behavior, for $|\alpha \Delta \Phi |\ll 1$,

\begin{equation}
V(\Phi) \simeq V_o (\alpha \, \Delta \Phi)^{\beta},  \label{asymp}
\end{equation}

\noindent and then, it behaves as an inverse power-law potential. Using eqs. (\ref{time},\ref{quad}), it can be demonstrated that this form of the potential is a solution when the scalar energy is strongly subdominant ($\rho^{\ast} \ll \rho_D$) with the same conditions given above for $F(a)$. In fact,

\begin{eqnarray}
\alpha \, \Delta \Phi \simeq a^{2 \frac{1+ \omega_{\Phi}}{%
\omega_{\Phi}-\omega_D}}  \label{aphi} \\
\Delta t \sim a^{\frac{3(1+\omega_D)}{2}}  \label{at} \\
\rho^{\ast} = \rho^{\ast}_o a^{-3(1+\omega^{\ast})}  \label{asym} \\
\omega^{\ast} = \frac{1+\frac{\beta}{2} \omega_D}{\frac{\beta}{2} -1}  \nonumber
\\
V^\prime (\Phi)= \beta \, \frac{V(\Phi)}{\Delta \Phi}  \label{afirst} \\
V^{\prime \prime} = \beta^2 \,\frac{V(\Phi)}{(\Delta \Phi)^2},
\label{asecond}
\end{eqnarray}

\noindent with $\rho _{o}^{\ast }$ given by eq.~\ref{rho}. In \cite{stein2},
it is demonstrated that an inverse power-law scalar potential has a
cosmological tracker solution, and then avoids the fine-tunning and the
cosmic coincidence problems of current cosmology. Therefore, our solution
is as follows: We will choose the appropriate values for the parameters in
the potential in such a way that the solution for eqs. (\ref{phi}-\ref{second}) can only be reached until the matter and the scalar energies are of the same order, as it happens today. From eqs. (\ref{alpha}-\ref{rho}), we can see that $\omega ^{\ast }$ and $\rho _{o}^{\ast }$ will become the current equation of state and the current energy density for the scalar field, respectively. Thus, all parameters can be determined and the solution becomes directly a tracker one. We can now draw a complete history for the Universe modeled by potential (\ref{potential}). The behavior of the scalar field for the completely radiation and matter dominated epochs is exactly the same than that found for an inverse power-law potential in \cite{stein2}, because at those epochs the scalar energy is subdominant. For instance, say that $\rho _{i\Phi }\leq \rho_{i\gamma }$ at the end of inflation. At the beginning, the kinetic energy dominates and $p_{\Phi }\simeq \rho _{\Phi }$, and eventually the evolution of the scalar energy stops, $p_{\Phi }\simeq -\rho _{\Phi }$, mimics a
``cosmological constant'' and remains frozen until the time when the tracker
solution is satisfied. However, the case $\rho _{i\Phi }>\rho
_{i\gamma }$ continues being troublesome. Another similar aspect is that our
potential reaches its tracker solution when $\Phi $ is of order $M_{P}$, as
can be seen from eq. (\ref{second}).

However, there are some key differences when the scalar field completely
dominates the evolution ($\rho _{\Phi }>\rho _{M}$). First, when passing
from a matter dominated epoch to a scalar field dominated one, the Universe
is driven naturally into a power-law inflationary stage\cite{chimen}.
Second, the scalar potential does not behave as a power-law potential
anymore. As it was found in\cite{urena}, the asymptotic behavior of the
scalar potential will be

\begin{eqnarray}
\Delta\Phi = \sqrt{3 \left( 1+\omega_{\Phi} \right)} \ln{a}  \label{SFD2} \\
V(\Phi) = \tilde{V_o} \, \exp{\ \left[- \sqrt{3 \left( 1+ \omega_{\Phi} \right)} \left( \Delta\Phi \right) \right]}  \label{SFD3} \\
\tilde{V_o} = \frac{V_o}{2^\beta} \nonumber \\
V^\prime = (\alpha \, \beta) \, V(\Phi) \\
V^{\prime \prime} = (\alpha \, \beta)^2 \, V(\Phi)
\end{eqnarray}

\noindent and then

\begin{equation}
H_o \Delta t = \frac{2}{3} \sqrt{\frac{12}{\Omega_{o\Phi}}} \, \frac{ a^{\frac{3%
}{2}(1+\omega_{\Phi})}}{(1+\omega_{\Phi})},  \label{SFD5}
\end{equation}

\noindent with $H_o$ the current Hubble parameter, and the scalar energy density will evolve as

\begin{equation}
\rho_\Phi = \frac{2}{1-\omega_\Phi} \, \tilde{V_o} \, a^{-3(1+\omega_\Phi)}. \label{solmd}
\end{equation}

\noindent Third, note that the equation of state $\omega _{\Phi }$ will
maintain the value it acquired in the matter dominated epoch, that is, it
will never switch to $-1$.

\begin{figure}[h]
\centerline{ \epsfysize=5cm \epsfbox{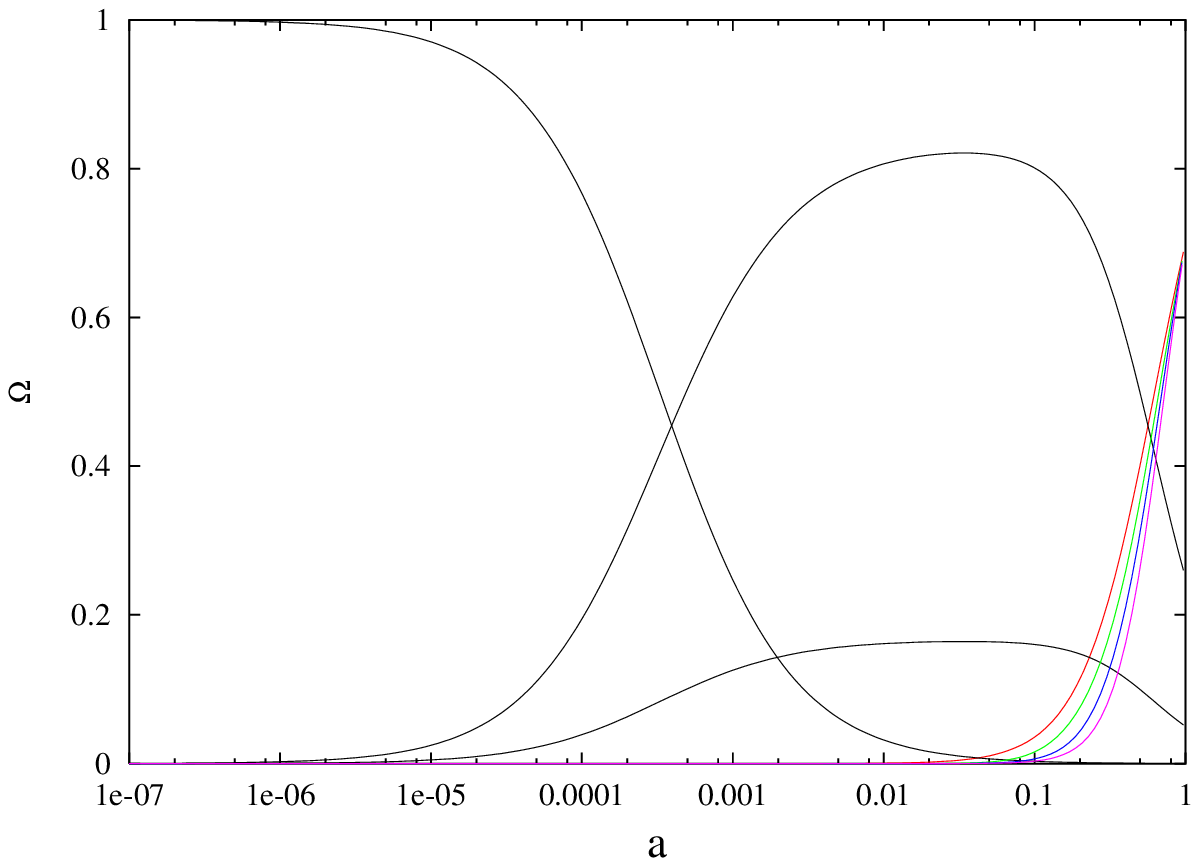}}
\label{fig:omegas}
\end{figure}

There is an almost generalized agreement in the current amounts for the
different components in the Universe\cite{perlmutter,triangle,stein}, thus,
we will take $\Omega _{oM}=0.30$ and $\Omega _{o\Phi }=0.70$. First of all,
in fig.~\ref{fig:omegas} it is shown the evolution of the density parameters
for each component of the Universe as functions of the scale factor $a$, for
the values $\omega _{\Phi }=\{-0.6,-0.7,-0.8,-0.9\}$. The scalar energy
remains subdominant until the epoch of matter domination and then the
tracker solution is reached. This must have happened at a redshift $6<z<19$.
Observe that a more negative $\omega $ causes a later scalar dominance. On
the other hand, the matter-scalar equality occurred at a redshift $0.3<z<0.6$%
.

\begin{figure}[h]
\centerline{ \epsfysize=5cm \epsfbox{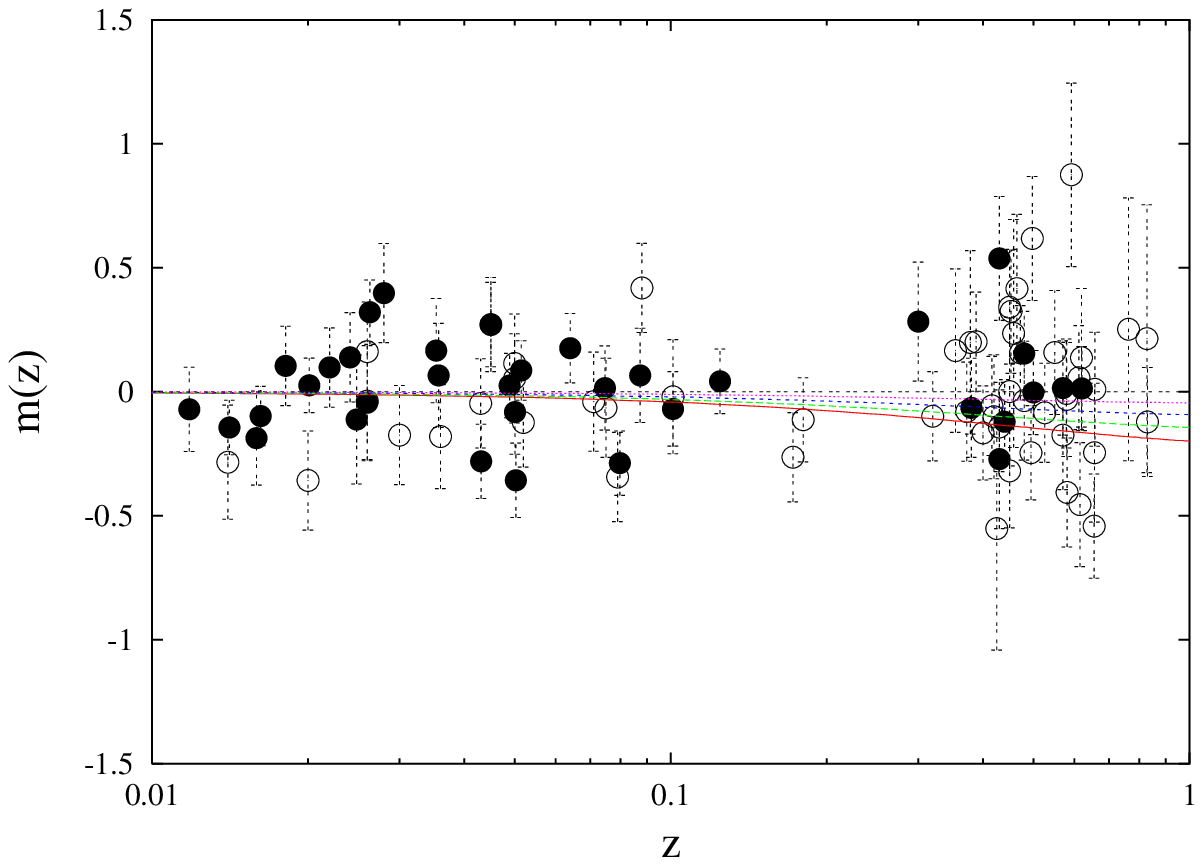}}
\label{fig:snia}
\end{figure}

Now, we will explore the consequences of the models for the different
observational constraints. Because the observations of SNIa has been done
for $z \leq 1$, it is easy to compute the luminosity distance with constant
scalar equation of state because the scalar field has already reached its
tracker solution. The results for different values of the equation of state
are shown in fig.~\ref{fig:snia}.

\begin{figure}[h]
\centerline{ \epsfysize=5cm \epsfbox{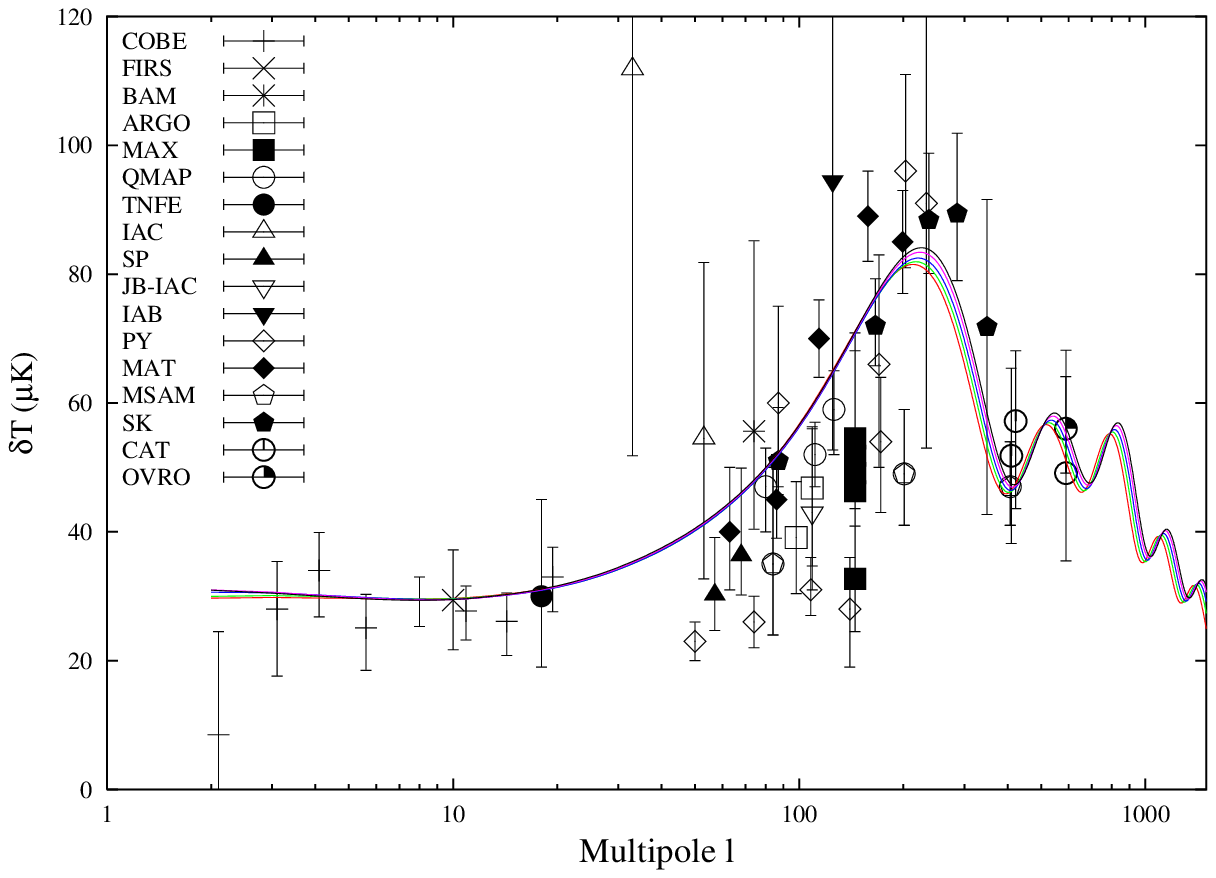}}
\label{fig:lamquint}
\end{figure}

We used an amended version of CMBFAST\cite{seljak} to compute the Angular
Power Spectrums for the different models in comparison with observational
data and $\Lambda CDM$ (fig. \ref{fig:lamquint}). The processed spectrums
are very similar.


\noindent Note that even for more negative $\omega_\Phi$ the more similar
the model is to the $\Lambda CDM$ one, there is no solution for the
potential (\ref{potential}) with $\omega_\Phi = -1$ (see eq. (\ref{alpha})).

\begin{figure}[h]
\centerline{ \epsfysize=5cm \epsfbox{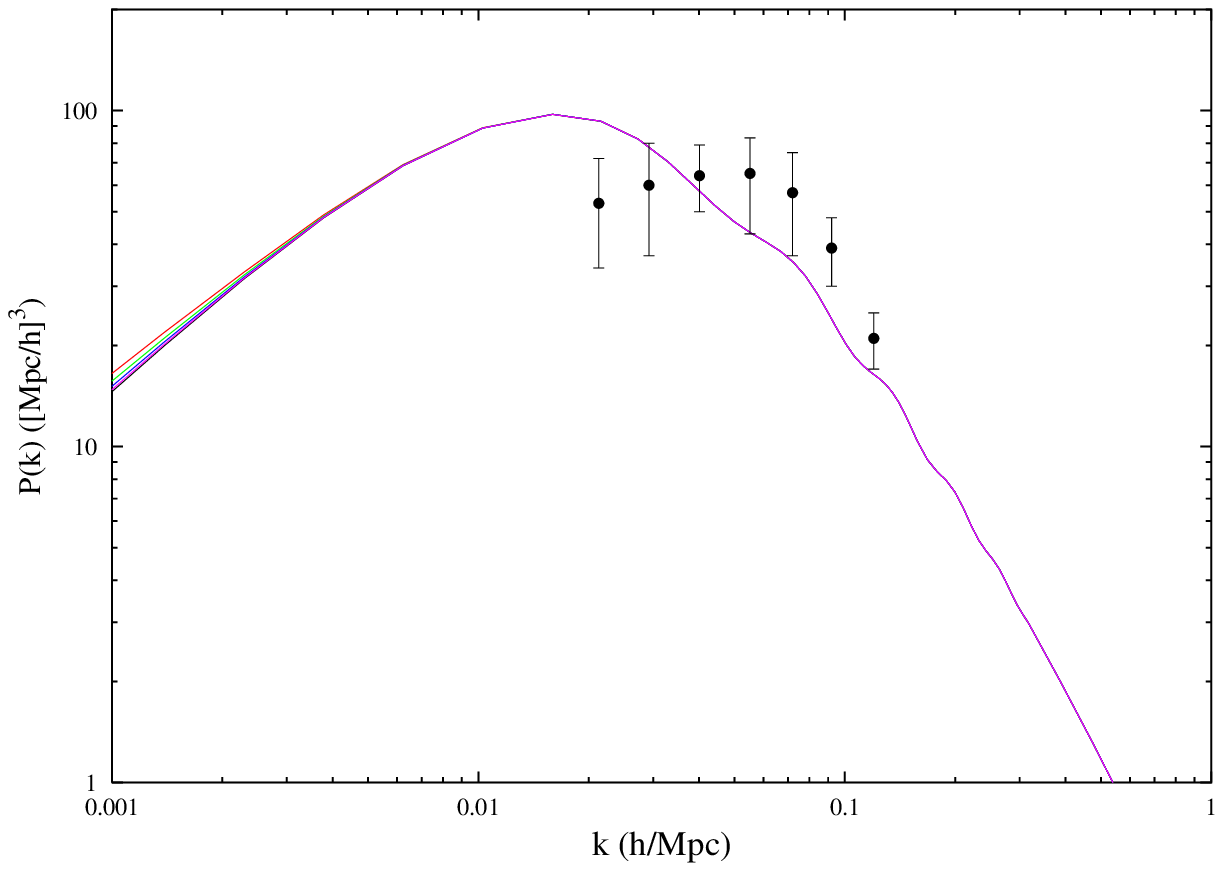}}
\label{fig:power}
\end{figure}

Following\cite{chung}, we also computed the Mass Power Spectrums (fig. \ref
{fig:power}).

\begin{figure}[h]
\centerline{ \epsfysize=5cm \epsfbox{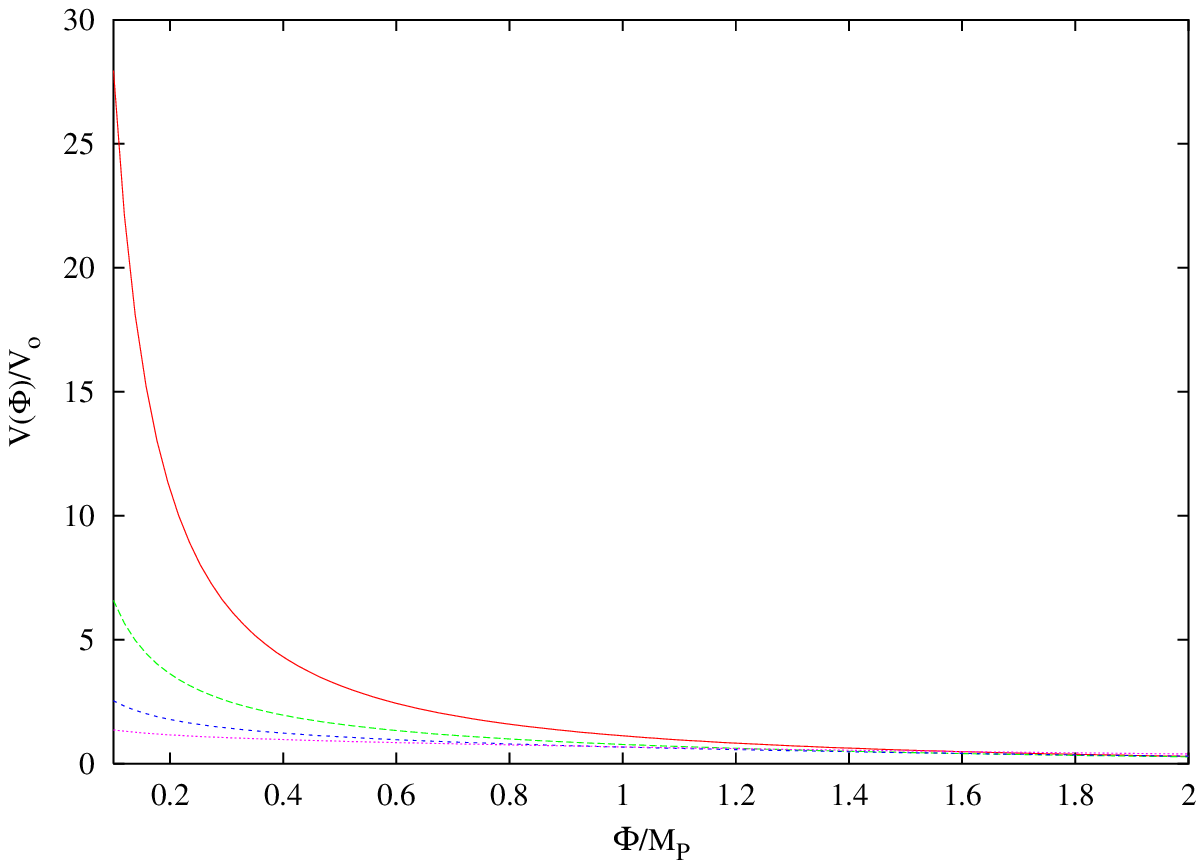}}
\label{fig:pot}
\end{figure}

\noindent We can observe the same similarity among the models, as it was
expected from the angular spectrum. As a reference, the potential is shown
in fig.~\ref{fig:pot} for different values of the equation of state (see
fig.~\ref{fig:omegas}).


The deceleration parameter for this model can be written in terms of the
present energy densities $\Omega _{oM}$, $\Omega _{o\Phi }$ and the scalar
equation of state $\omega _{\Phi }$

\begin{equation}
q_o \equiv - H^2_0 ( \ddot{a}/a)_0 = \frac{1}{2} \Omega_{oM} + \frac{1+3
\omega_{\Phi}}{2} \Omega_{o\Phi}.
\end{equation}

\noindent Thus, $-0.13 > q_o > -0.4 $. The age of the Universe is $ ( 13 < t_o
 < 14 ) \times 10^{9}$ years.

Summarizing, we have found that the potential

\begin{eqnarray}
V(\Phi) = \frac{1-\omega_\Phi}{2} \, \rho_{o\Phi} \, \left( \frac{\Omega_{oM}}{
\,\Omega_{o\Phi}} \right)^{\frac{1+\omega_\Phi}{\omega_\Phi}}  \times \nonumber \\
\left[ \sinh{\left( \frac{-3\omega_\Phi}{2 \sqrt{3(1+\omega_\Phi)}} \,
\Delta \Phi \right)} \right]^{\frac{2(1+\omega_\Phi)}{\omega_\Phi}}
\end{eqnarray}

\noindent is a good quintessential candidate to be the missing energy in the
Universe. Its behavior as an inverse power-law potential at early times
allows us to avoid the fine tuning problem under the initial condition (say,
at the end of inflation) $\rho _{i\Phi }<\rho _{i\gamma }$. Its exponential
like-behavior would drive the Universe into a power-law inflationary stage,
which is in accord with recent observations. The three parameters ($V_o$, $\alpha$ and $\beta$ in (\ref{potential})) can be determined uniquely by the measured values for the equation of state and the amount of vacuum energy to obtain a tracker solution, then we can avoid the coincidence problem. Also, we found a good agreement with the current observations for SNIa, Angular and Mass power spectrums. However, we do not already know about a theory that could predict this kind of potentials, but we think that its attractive features make it worth to study.

\section{Acknowledgements}

This work was partly supported by CONACyT, M\'{e}xico 119259 (L.A.U.)



\newpage

\section{captions}

FIG. 1. Evolution of the dimensionless density parameters taking $%
\Omega_{oM} = 0.30$ and $\Omega_{o\Phi}=0.70$. The different models with $%
\protect\omega_{\Phi} = \{-0.6 (red),-0.7 (green),-0.8 (blue),-0.9 (pink) \}$
are also shown.

FIG. 2. The luminosity distance {\it vs} redshift for the models shown in
fig.~\ref{fig:omegas}. The horizontal line is the luminosity distance for $%
\Lambda CDM$. The open circles represent the observational results from SCP
and the opaque circles are from HZS\protect\cite{perlmutter}.

FIG. 3. Angular power spectrum for the models considered in fig.~\ref
{fig:omegas} and $\Lambda CDM$ (black). The data were taken from\protect\cite
{belen}.

FIG. 4. Mass power spectrum for the models shown in fig.~\ref{fig:omegas}.
All models with spectral index $n_s=1$. The data points are from\cite{peacock}.

FIG. 5. Evolution of the scalar potential $V(\Phi)$ in terms of the scalar
field $\Phi$.

\end{document}